\documentclass[12pt]{article}
\usepackage{amsmath,amssymb,amsfonts}
\usepackage{graphicx}
\usepackage{verbatim}

\newcommand{\figref}[1]{Fig. \ref{#1}}

\usepackage{hyperref} 

\newcommand{\lt}{\left}
\newcommand{\rt}{\right}

\newcommand{\bit}{\begin{itemize}}
\newcommand{\eit}{\end{itemize}}
\newcommand{\ben}{\begin{equation}}
\newcommand{\een}{\end{equation}}
\newcommand{\bea}{\begin{eqnarray}}
\newcommand{\eea}{\end{eqnarray}}
\newcommand{\bens}{\begin{equation*}}
\newcommand{\eens}{\end{equation*}}
\newcommand{\beas}{\begin{eqnarray*}}
\newcommand{\eeas}{\end{eqnarray*}}

\newcommand{\half}{{1\over2}}

\newcommand{\foot}{\footnote}

\newdimen\addresswidth
\numberwithin{equation}{section}

\begin{document}

\begin{titlepage}
\hfill
\vbox{
    \halign{#\hfil         \cr
           } 
      }  
\vspace*{20mm}
\begin{center}
{\Large \bf  Flavor-symmetry Breaking with Charged Probes }

\vspace*{15mm}
\vspace*{1mm}
{Joshua L. Davis,\footnote{\tt e-mail: jdavis@phas.ubc.ca} and Namshik Kim\footnote{\tt e-mail: namshik@phas.ubc.ca}}
\vspace*{1cm}

{\it Department of Physics and Astronomy,
University of British Columbia\\
6224 Agricultural Road,
Vancouver, B.C., V6T 1W9, Canada}

\vspace*{1cm}
\end{center}

\begin{abstract}

We discuss the recombination of brane/anti-brane pairs carrying $D3$ brane charge in $AdS_5 \times S^5$. These configurations are dual to co-dimension one defects in the ${\cal N}=4$ super-Yang-Mills description. Due to their $D3$ charge, these defects are actually domain walls in the dual gauge theory, interpolating between vacua of different gauge symmetry. A pair of unjoined defects each carry localized $(2+1)$ dimensional fermions and possess a global $U(N)\times U(N)$ flavor symmetry while the recombined brane/anti-brane pairs exhibit only a diagonal $U(N)$. We study the thermodynamics of this flavor-symmetry breaking  under the influence of external magnetic field.

\end{abstract}

\setcounter{footnote}{0}
\end{titlepage}

\section{Introduction}

The $AdS/CFT$ correspondence \cite{Maldacena:1997re}, and holographic duality in general, is a powerful, conjectured technique for the analysis of strongly coupled field theories. While originally pursued to address questions about low-energy QCD, it has expanded to include studies of a variety of strongly coupled field theories in diverse dimensions.\foot{For older review articles see \cite{Aharony:1999ti,D'Hoker:2002aw}, while \cite{Hartnoll:2009sz,McGreevy:2009xe} are more recent with an emphasis on applications for condensed matter.}

Of much interest in recent years has been the study of defect theories and the interaction of defects. Such defects can be constructed holographically by the intersection of different stacks of $D$-branes, one of the earliest known examples being the supersymmetric $(2+1)$-dimensional intersection of the $D3/D5$ system \cite{DeWolfe:2001pq}, a defect in the ambient $(3+1)$-dimensional ${\cal N}=4$ super Yang-Mills native to the $D3$ worldvolume. A common technique for studying these systems is to consider the quenched approximation of the field theory, where one stack, say of $Dp$-branes, has parametrically more branes than the other, say of $Dq$-branes. The gravity description of this scenario can then be reliably computed at strong coupling by using a probe $Dq$-brane action in the near-horizon region of a classical $p$-brane supergravity solution \cite{Karch:2002sh}. The full dual field theory lives at the asymptotic boundary of this spacetime and the defect theory lives where the probe brane intersects the boundary.

Multiple defects may be studied by allowing several stacks of $Dq$-branes to intersect the boundary. As discussed first in \cite{Sakai:2004cn,Sakai:2005yt}, a coherent state of spatially separated defects can be achieved by a continuous probe brane configuration with a multiply connected intersection with the boundary. Since the boundary components must have opposite orientation in this scenario, it can be understood as brane/anti-brane recombination. In the scenario of \cite{Sakai:2004cn,Sakai:2005yt}, the defect degrees of freedom were $d=3+1$ chiral fermions, with those on the brane component of opposite chirality from those on the anti-brane. The coherent state where the worldvolumes join in the bulk thus describes chiral symmetry breaking. In \cite{Antonyan:2006vw,Antonyan:2006pg}, this scenario was generalized to allow for intersections of other dimension and brane species as well as for the joining process to occur dynamically.\foot{In \cite{Sakai:2004cn,Sakai:2005yt}, topological considerations force the branes to join while in \cite{Antonyan:2006vw,Antonyan:2006pg} and later works there are multiple consistent solutions and only the minimum energy one dominates.} Further generalizations have included adding external magnetic and electric fields as well as chemical potential \cite{Parnachev:2006ev,Davis:2007ka,Bergman:2008sg,Johnson:2008vna,Johnson:2009}.

In this paper, we consider scenarios of bulk brane/anti-brane recombination in $AdS_5 \times S^5$, 
\ben
ds^2\sim r^2 \lt(-dt^2 + dx^2 + dy^2 + dz^2\rt) + {dr^2\over r^2} + d\Omega_5^2~.
\een
As an additional ingredient to previous studies, we consider probes which are electrically charged under the background $F_5$ Ramond-Ramond field. The probe branes form two stacks, each spanning some cycle in $S^5$, the non-compact directions $(t,x,y)$ and some curve $z(r)$. The stacks have opposite orientation and are separated in the $z$ direction along the boundary.

\begin{figure}
\centering
\begin{tabular}{cc}
\includegraphics[width=.5\textwidth]{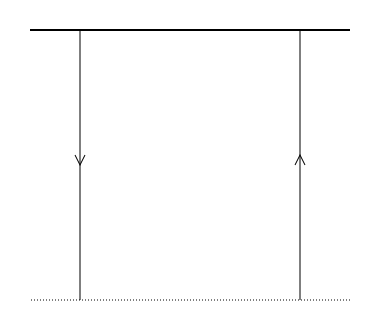} &
\includegraphics[width=.5\textwidth]{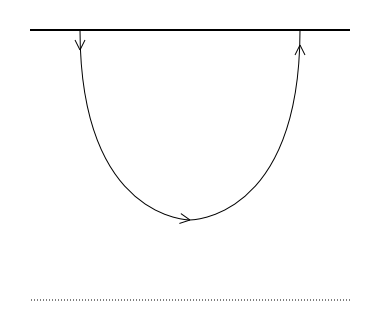}
\end{tabular}
\caption{\label{uncharged} Straight embeddings and a joined embedding where there is no force from the background Ramond-Ramond flux. The arrows represent worldvolume orientation. There would be no change in the embedding if the arrows were reversed.}
\end{figure}

An uncharged probe brane -- such as in the studies cited above -- experiences no force in the non-compact directions from the $F_{5}$. For such a case there are then two qualitative classes of solutions, depicted in \figref{uncharged}. The first solution is the so-called ``black hole embedding'' which reaches all the way down to the spacetime horizon. These embeddings are ``straight'' in the sense that ${dz \over dr}=0$. The second solution is a joined embedding which has two disconnected boundaries of opposite orientation although the entire worldvolume is a simply connected and oriented manifold. Only these solutions have ${dz \over dr}\ne0$. Note that since there is no Ramond-Ramond force, the brane orientation does not play a role.

\begin{figure}
\centering
\includegraphics[width=.6\textwidth]{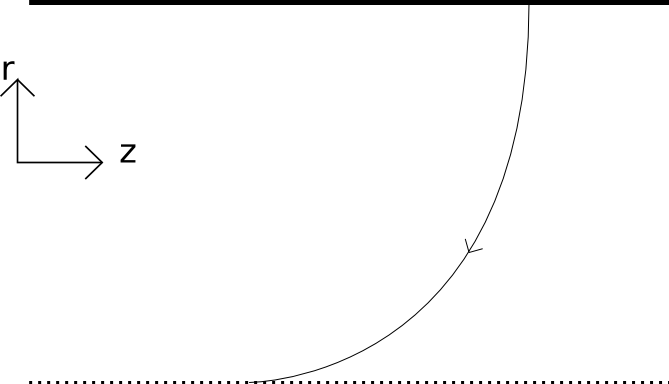} 
\caption{\label{chargedprobe} A $D3$-charged probe brane in (finite-temperature) $AdS_5 \times S^5$. The probe bends in the $z$-direction as it descends from the boundary (the solid line at the top) to enter the horizon represented by the dotted line at the bottom. The arrow represents the orientation of the $D3$ charge. An oppositely oriented brane would bend in the opposite $z$-direction.}
\end{figure}

On the other hand, if the probe branes are charged under the spacetime Ramond-Ramond field, the situation is somewhat different. This can occur either because the probe itself is a $D3$-brane, or the charge could be induced by worldvolume fluxes on the probe. The $D3/D5$ system where the $D5$ brane carries $q$ unit of $D3$-brane charge 
was first introduced in \cite{Karch:2001}. In  \cite{Myers:2008me}, black-hole embeddings of $D5$ and $D7$ probe branes with induced $D3$-brane charge were studied in $AdS_5\times S^5$. Additional $D7$ brane embeddings carrying $D3$ charge were introduced in \cite{Bergman:2010gm} and studied further in \cite{Jokela:2010nu,Davis:2011gi}. These probes are affected by the background $F_5$ and even the black hole embeddings have ${dz\over dr}\ne0$. In \figref{chargedprobe}, we see such a black-hole embedding. The brane orientation plays a major role in this situation; an oppositely oriented probe would bend in the opposite $z$-direction.

\begin{figure}
\centering
\begin{tabular}{cc}
\includegraphics[width=.5\textwidth]{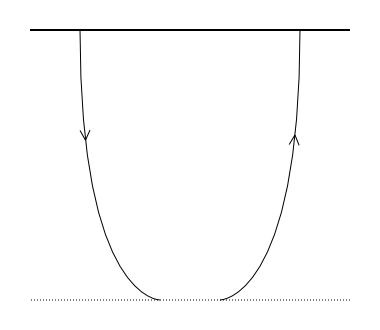} &
\includegraphics[width=.5\textwidth]{skinny.png} \\
\includegraphics[width=.5\textwidth]{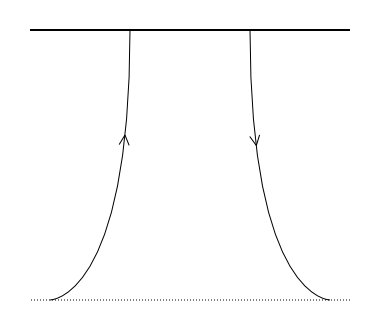} &
\includegraphics[width=.5\textwidth]{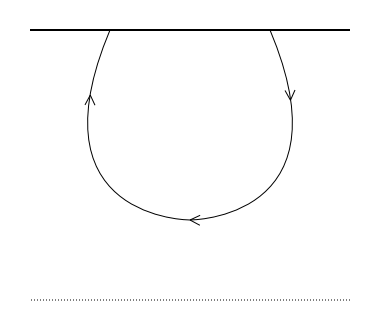} \\
\end{tabular}
\caption{\label{chargedjoin} The top row pictures possible solutions of a brane/anti-brane pair in the presence of a Ramond-Ramond force. Note that the branes bend toward each other as they extend into the bulk even if they don't join. If the orientations are reversed, we have instead the bottom set of solutions. These always bend away from each other when initially leaving the boundary even if they eventually join in the bulk.}
\end{figure}

These electrically charged probe branes have a richer space of joined solutions than their uncharged cousins. Due to the force in the $z$-direction, the qualitative features of the solution depend strongly on the orientation, specifically the left-right ordering of the boundary components. The choice of orientation gives rise to the classes of solutions seen in \figref{chargedjoin}. The top left figure pictures a brane/anti-brane pair which tend toward each other despite not actually connecting, while the top right figure pictures a joined pair. These two solutions have the same boundary conditions and so it is a dynamical question which has the lower energy and is therefore stable. The figures in the bottom row also depict solutions with the same boundary conditions, but with the worldvolume orientations all opposite of the figures above. Note the surprising feature in the bottom right figure, where the joined embedding becomes wider in the bulk than at the boundary. We will call these joined solutions ``chubby'' and conversely the more typical solutions in the top right (which are widest at the boundary) we will call ``skinny.''

There are multiple perspectives on what these brane systems are holographically dual to. Firstly, the $(2+1)$-dimensional intersection of a probe brane with the boundary is conventionally associated with a defect in ${\cal N}=4$ super-Yang-Mills gauge theory. The field content of the defect is given by the lowest level open string modes which are localized at the $D$-brane intersection. For a $D5$-brane probe, the defect theory is supersymmetric  since the intersection is $\#ND=4$; this is the spectrum studied in \cite{DeWolfe:2001pq}. For the $D7$-brane probe, the intersection is $\#ND=6$ and the spectrum is simply massless fermions \cite{Polchinski:1998rr}, in fact T-dual to the $D4/D8$ intersections of the Sakai-Sugimoto model \cite{Sakai:2004cn,Sakai:2005yt}. As a caveat, it should be mentioned that it is not clear if this picture of the spectrum still holds when internal fluxes exist on the probe, but is often nonetheless used to guide intuition.

A defect dual to a stack of $N$ branes or anti-branes is associated with a $U(N)$ global flavor symmetry inherited from the gauge field living on the brane worldvolume. Thus the recombination of an equal number of branes and anti-branes describes a breaking of symmetry $U(N)\times U(N)\to U(N)$. Since the defects are separated in space, the duals of these scenarios can be considered interacting $(2+1)$-dimensional defect bi-layer systems or as discussed in \cite{Antonyan:2006pg}, $(2+1)$-dimensional effective field theories with non-local interactions.

\begin{figure}[t]
\begin{center}
\includegraphics[width=.6\textwidth]{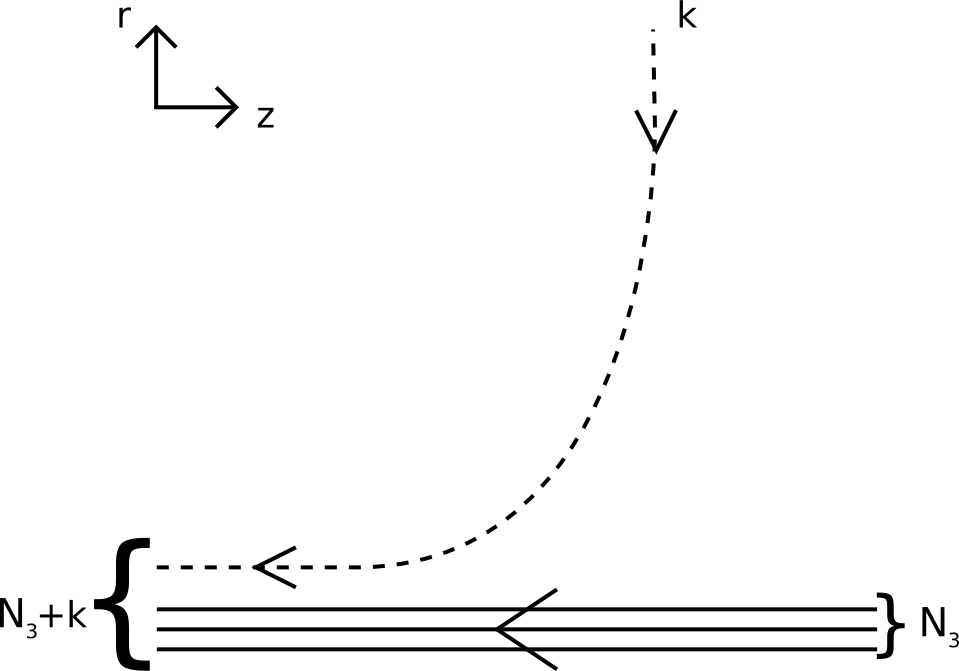}
\end{center}
\caption{\label{basiccartoon} A cartoon representation of a probe brane (dashed line) carrying $D3$-charge $k$ bending to become parallel with the stack of $N_3$ $D3$-branes sourcing the $AdS$ geometry, represented by the solid lines at the bottom. The arrows represent brane worldvolume orientation. The dual gauge group is $SU(N_3)$ towards the right while it is enhanced to $SU(N_3+k)$ to the left. }
\end{figure}

The dual interpretation above holds for probes with or without $D3$-charge. However, for $D3$-charged probes there are some other interesting properties of these solutions. A $D3$-charged probe brane -- even a higher dimensional brane with an induced $D3$ charge -- contributes to the overall Ramond-Ramond flux of the system. This flux is in turn related by the $AdS/CFT$ dictionary to the rank of the dual gauge group. Therefore a defect of $D3$-charge $k$ forms a domain wall in the dual gauge theory with $SU(N)$ gauge symmetry on one side and $SU(N+k)$ on the other \cite{Myers:2008me}. A cartoon representation of this situation is depicted in \figref{basiccartoon}. Once the probe enters the horizon, it is effectively parallel to the original stack of $D3$-branes sourcing the geometry, adding to the overall $D3$-brane charge as measured by a Gaussian surface outside the horizon. This is interpreted as a larger gauge symmetry existing in the region to the left. It follows that if there are multiple $D3$-charged defects, that we have a spatially non-trivial pattern of symmetry breaking in the dual theory, with a gauge group between the defects which is different from that outside. Thus the joined solutions should be considered dual to finite-width domain walls.

In this paper, we will study the thermodynamics of these domain walls, mostly from the bulk perspective. In Section 2, we introduce a class of $D3$-charged probe branes and derive a one-dimensional effective particle mechanics action that describes the entire class. The solutions of the equation of motion of this effective action are studied in Section 3 and a renormalized free energy computed in Section 4. Finally, in Section 5, we examine the phase diagram of this system in the space of external magnetic field and asymptotic separation with some comments on the phenomenon of magnetic catalysis.

\section{$D3$-charged probes in $AdS_5 \times S^5$}

Consider the background $IIB$ supergravity solution thermal $AdS_5 \times S^5$, the near-horizon geometry of $N_3$ $D3$-branes at finite temperature. The line-element is given by
\ben\label{ads5s5}
L^{-2} ds^2 = r^2\lt(-h(r)dt^2 + dx^2+dy^2 +dz^2\rt) +{dr^2 \over h(r) r^2} + d\Omega_5^2~.
\een
The $S^5$ line element is represented as a bundle over $S^2\times S^2$,
\ben
d\Omega_5^2 = d\psi^2 + \sin^2 \psi d\Omega_2^2 + \cos^2\psi d{\tilde{\Omega}_2^2}~,
\een
where $\psi\in \lt(0 ,{\pi\over2}\rt)$ and
\ben
d\Omega_2^2=d\theta^2 + \sin^2 \theta d\phi^2~,
\een 
is the line-element for a unit $S^2$. The blackening function is
\ben
h(r) =1-{r_h^4\over r^4}~.
\een
At zero-temperature, $r_h=0$. However, any non-zero value of $r_h$ can be rescaled by a coordinate transformation. Therefore, for finite temperature, we can choose without loss of generality $r_h=1$. The scale of the geometry is related to the microscopic string theory parameters via
\ben
L^4 = 4\pi g_s N_3 (\alpha^\prime)^2~.
\een
There is also a self-dual five-form Ramond-Ramond flux
\ben
F_5 =  {4 L^4\over g_s}\lt(r^3 dt \wedge dx \wedge dy\wedge dz\wedge dr + \omega_5\rt)~.
\een
Here $\omega_5$ is the volume form on the unit five-sphere
\ben
\omega_5 = \sin^2 \psi \cos^2 \psi d\psi \wedge \omega_2 \wedge \tilde{\omega}_2~,
\een
where $\omega_2=\sin\theta d\theta\wedge d\phi$ is the $S^2$ volume form. We encode this flux with the four-form potential
\ben\label{c4}
{g_s\over L^4}C_4 =  r^4 h(r) dt\wedge dx\wedge dy\wedge dz + \half c\lt(\psi\rt) \omega_2 \wedge {\tilde \omega}_2~.
\een
The function $c\lt(\psi\rt)$ is
\ben
c(\psi) = \psi -{1\over4}\sin\lt(4\psi\rt) + c_0
\een
where $c_0$ is an arbitrary constant, a residual ambiguity due to the gauge symmetry of the Ramond-Ramond field. A similar constant could be added to the coefficient of $dt\wedge dx\wedge dy \wedge dz$, but we have chosen to partially fix the gauge by requiring that the first term in $C_4$ vanish at the horizon. This ensures that the term is well-defined on the Euclidean section of \eqref{ads5s5} which simplifies the treatment of the Wess-Zumino terms.

We will now consider the following set of branes
\ben\label{dbranes}
\begin{array}{rcccccccccccl}
  & & t & x & y & z & r & \psi & \Omega_2 & \tilde{\Omega}_2 & \\
& D3'& - & - & - & - & \cdot & \cdot &  \cdot & \cdot& \\
& D3 & - & - & - & \sim & \sim & \cdot &  \cdot & \cdot& \\
& D5 & - & - & - & \sim & \sim & \cdot &  - & \cdot& \\
& D7 & - & - & - & \sim & \sim & \cdot &  - & -& \\
\end{array}
\een
The $D3'$ row refers to the large stack of $D3$ branes which source the $AdS$ geometry while the other rows record the configurations of the probes. A dash indicates the brane is extended in that direction, with support over the entire range of the coordinate. A dot indicates the respective brane is completely localized in that coordinate. Finally the $\sim$ symbols indicate that the brane traces a curve in those directions. For example, the $D5$-brane extends along the non-compact $(t,x,y)$ directions, wraps one of the two $S^2$ factors in the $S^5$, is localized in $\psi$ and on the other $S^2$, and finally, lies along a curve in the $(z,r)$ space. 

These probes all intersect the boundary on some $2+1$ dimensional subspace at a fixed value of $z$ (although for the $D3$ probes, this will turn out to be $z=\pm\infty$). In order to induce $D3$-brane charge,\foot{Such flux is actually required to stabilize the $D7$ probe at a non-trivial value of $\psi$ at the $AdS$ boundary \cite{Bergman:2010gm}.} the probe $D5$ and $D7$-branes will carry internal flux topologically supported on one or both $S^2$ factors, respectively. We will also allow for magnetic field in the three dimensional defect on the boundary, {\it i.e.} a non-zero $F_{xy}$ component. In \cite{Bergman:2010gm}, $D7$ branes with a more general {\it ansatz} were studied. However, our focus will be a class of solutions with different boundary conditions.

The $D5$ and $D7$ probes outlined above have $3+1$ non-compact directions and wrap some compact cycles. If one imagines integrating over these cycles, one would obtain an effective $3+1$ dimensional object which carries $D3$ charge in $AdS_5$. In other words, the higher-dimensional $D3$-charged branes act as effective $D3$-branes. These effective branes are much like excited states of a proper $D3$, they carry $D3$ charge but the effective tension is greater than the charge. This will become clearer in the next few sections. First, we will calculate an effective action for a $D3$ probe with the {\it ansatz} \eqref{dbranes}. We will then see that $D5$ and $D7$ probes will yield an effective action of the same form.

\subsection{$D3$-brane probe}

First, let us introduce a $D3$-brane probe as a model system. The action comprises the familiar DBI and Wess-Zumino terms
\ben
S_3= -T_3 \int d^{3+1} \xi e^{-\phi} \sqrt{-{\rm det}\lt(g+2\pi\alpha' F\rt)} -  T_3 \int C_4~.
\een
The three-brane tension is
\ben
T_3 = {1\over (2\pi)^3}{1\over {\alpha'}^2}~.
\een
We choose a static gauge where $\xi^a=\lt\{t,x,y,r\rt\}$ are brane coordinates and the embedding is given by the function $z(r)$. The induced metric is thus
\ben
{ds_3^2 \over L^2} = r^2\lt(-hdt^2+dx^2+dy^2\rt) + \lt(1+ r^4 h \dot{z}^2\rt) {dr^2\over r^2 h}~,
\een
where a dot indicates differentiation by $r$. We also allow a magnetic field normalized as
\ben
{2\pi \alpha'\over L^2} F = B dx \wedge dy~.
\een
This information is sufficient to compute the Born-Infeld term 
\ben
S_{DBI}= {\cal N}_3 \int dr \sqrt{\lt(r^4+B^2\rt)\lt(1+ r^4 h \dot{z}^2\rt)}~,
\een
where the overall constant is
\ben
{\cal N}_3={T_3 L^4 V_{2+1} \over g_s}~,
\een
with $V_{2+1}$ the infinite volume factor of the $(t,x,y)$ directions.

To compute the Wess-Zumino term we also need to specify an orientation, which we encode via an orientation parameter $\zeta=\pm1$. Evaluating,
\ben
\int C_4 = V_{2+1} \zeta \int r^4 h \dot{z} dr~.
\een
Note that while orientation is an invariant geometric feature intrinsic to the entire $D$-brane worldvolume, the parameter $\zeta$ is partly an artifact of the coordinates we use. Therefore $\zeta$ may take different values on separate branches of the same continuous brane. For example, in a brane/anti-brane recombination, the left branch has $\zeta=1$ and the right branch $\zeta=-1$, yet the worldvolume is continuous.

Putting together the terms above -- and dropping an overall constant factor -- yields an effective particle mechanics Lagrangian
\ben
L_3 =  \sqrt{\lt(r^4+B^2\rt)\lt(1+ r^4 h \dot{z}^2\rt)} + \zeta r^4 h \dot{z}~.
\een
We will find similar effective Lagrangians for the $D5$ and $D7$ probes, differing only in the coefficient of the second term. Here that coefficient is of unit magnitude since physically it is the $D3$-brane charge per tension.

\subsection{$D5$ probes}

The probe action for $D5$-branes is
\ben\label{d5action}
S_5 = - T_5 \int d^{5+1} \xi e^{-\phi} \sqrt{-{\rm det}\lt(g+2\pi\alpha' F\rt)} - 2\pi\alpha' T_5 \int C_4 \wedge F~,
\een
where the tension is
\ben
T_5 = {1\over (2\pi)^5}{1\over {\alpha'}^3}~.
\een
We choose a static gauge with coordinates $\xi^a=\lt\{t,x,y,r, \theta,\phi\rt\}$ and embedding function $z(r)$. The induced metric is
\ben
{ds_5^2 \over L^2} = r^2\lt(-hdt^2+dx^2+dy^2\rt) + \lt(1+ r^4 h \dot{z}^2\rt) {dr^2\over r^2 h} + \sin^2\psi d\Omega^2_2~.
\een
The {\it ansatz} for worldvolume flux is
\ben
{2\pi \alpha'\over L^2} F = B dx \wedge dy + {f \over 2} \omega_2 ~.
\een
The magnetic field is a continuous quantity but the flux on the compact sphere is, of course, quantized
\ben
f = {2\pi \alpha' \over L^2} n~, \quad\quad\quad n\in\mathbb{Z}~.
\een

Substituting all this into the action yields
\ben
{S_5}=- {\cal N}_5 \int dr\lt( \sqrt{\lt(r^4+B^2\rt)\lt(f^2 + 4 \sin^4\psi\rt) \lt(1+ r^4 h \dot{z}^2\rt)} + \zeta f r^4 h \dot{z}\rt)~,
\een
with the normalization
\ben
{\cal N}_5={2\pi T_5 L^6 V_{2+1} \over g_s}~,
\een 
and once again we have introduced an orientation parameter $\zeta=\pm1$. Our {\it ansatz} is for constant $\psi$ but we see that $\psi$ has a potential. The $\psi$ equation of motion is
\ben
{d~\over d\psi} \sqrt{f^2+4\sin^4\psi}=0~,
\een
yielding\foot{Another solution is $\psi=0$ but it is physically trivial since the brane volume is then exactly zero.}
\ben
\psi = {\pi \over 2}~.
\een
We insert this back into the $D5$ action. Up to an overall constant we again obtain an effective particle Lagrangian for $z(r)$,
\ben
L_5 =  \sqrt{\lt(r^4+B^2\rt)\lt(1+ r^4 h \dot{z}^2\rt)} + {\zeta f\over \sqrt{f^2 +4}} r^4 h \dot{z}~.
\een
The only difference from the $D3$ is in the coefficient of the second term, the effective $D3$-brane charge per unit tension. The magnitude of this ratio is less than unity here, in keeping with the picture that this $D5$ probe is a $D3$-brane in an excited state.

\subsection{$D7$ probes}

The $D7$-brane action is
\ben\label{d7action}
S_7 = - T_7 \int d^{7+1} \xi e^{-\phi} \sqrt{-{\rm det}\lt(g+2\pi\alpha' F\rt)} - {\lt(2\pi\alpha'\rt)^2\over 2} T_7 \int C_4 \wedge F\wedge F~,
\een
with tension
\ben
T_7 = {1\over (2\pi)^7}{1\over {\alpha'}^4}~.
\een
In a static gauge with coordinates $\xi^a=\lt\{t,x,y,r, \theta,\phi, \tilde{\theta}, \tilde{\phi}\rt\}$, we describe the embedding with the function $z(r)$. The induced metric is
\ben
{ds_7^2 \over L^2} = r^2\lt(-hdt^2+dx^2+dy^2\rt) + \lt(1 + r^4 h \dot{z}^2\rt) {dr^2\over r^2 h} + \sin^2\psi d\Omega^2_2+\cos^2\psi d\tilde{\Omega}^2_2~.
\een
For the worldvolume flux we use the {\it ansatz}
\ben
{2\pi \alpha'\over L^2} F = B dx \wedge dy + {f_1 \over 2} \omega_2 + {f_2 \over 2} \tilde{\omega}_2~.
\een
The fluxes on the $S^2$ factors are quantized
\ben
f_i = {2\pi \alpha' \over L^2} n_i~, \quad\quad\quad n_i\in\mathbb{Z}~.
\een

The DBI portion of the action is
\ben
S_{DBI} = -{\cal N}_7\int dr \sqrt{\lt(r^4+B^2\rt)\lt(f_1^2+4\sin^4{\psi}\rt)\lt(f_2^2+4\cos^4{\psi}\rt)\lt(1 +r^4 h \dot{z}^2\rt)}
\een
with
\ben
{\cal N}_7 = {4\pi^2 T_7 L^8 V_{2,1}\over g_s}~.
\een
The Wess-Zumino term is given by
\ben
S =  -{\cal N}_7 \zeta f_1 f_2 \int dr r^4 h \dot{z}~,
\een
with $\zeta$ the orientation parameter. We minimize the $\psi$ potential
\ben
{d~\over d\psi} \sqrt{\lt(f_1^2+4\sin^4\psi\rt)\lt(f_2^2+4\cos^4\psi\rt)}=0~,
\een
yielding the implicit equation\foot{While this can be solved for general $f_i$, it can be seen that fluctuations $\delta \psi$ around the solution can violate the BF bound \cite{Breitenlohner:1982bm,Breitenlohner:1982jf}. In particular, for absolutely no internal fluxes $f_i=0$, the $D7$ will be unstable \cite{Davis:2008nv}. See \cite{Bergman:2010gm} for more discussion of stabilizing this $D7$ brane embedding.}
\ben\label{psisol}
f_2^2 \sin^2 \psi -f_1^2 \cos^2 \psi +4 \cos^2 \psi \sin^2\psi \lt(\cos^2\psi-\sin^2\psi\rt)=0~.
\een
Substituting this back into the action yields, up to an overall constant, an effective particle Lagrangian for the $D7$-brane
\ben
L_7 =  \sqrt{\lt(r^4+B^2\rt)\lt(1+ r^4 h \dot{z}^2\rt)} + {\zeta f_1 f_2\over \sqrt{\lt(f_1^2 +4\sin^4\psi_0\rt)\lt(f_2^2+4\cos^4\psi_0\rt)}} r^4 h \dot{z}~,
\een
where $\psi_0$ is a constant that solves \eqref{psisol}. This again takes the form of the effective $D3$ Lagrangian with a charge per tension smaller than unity.

\section{Solutions to effective Lagrangian}

We found that all three of the $D3$-charged probes under consideration are described by an effective particle Lagrangian of the form
\ben\label{efflag}
S_{eff}= \int dr \sqrt{r^4+B^2} \sqrt{1+ r^4 h \dot{z}^2} +\alpha \int dr r^4 h \dot{z}~.
\een
The parameter $\alpha$ is the effective $D3$-brane charge per tension and is given by
\ben\label{alphacases}
\alpha =
\begin{cases}
\zeta & D3{\rm-brane}\\
{\zeta f \over \sqrt{f^2 +4}} & D5{\rm-brane}\\
{\zeta f_1 f_2 \over \sqrt{\lt(f_1^2 +4\sin^4\psi_0\rt)\lt(f_2^2+4\cos^4\psi_0\rt)}} & D7{\rm-brane}
\end{cases}
\een
with $\psi_0$ solving \eqref{psisol} in the case of the $D7$. Note that $|\alpha|<1$ for both the $D5$ and $D7$ probes.

The equation of motion derived from \eqref{efflag} can be immediately integrated since the variable $z(r)$ is cyclic
\ben\label{pz}
P \equiv {\sqrt{r^4 +B^2\over1+r^4 h \dot{z}^2} }r^4 h \dot{z}+\alpha r^4 h = constant~.
\een
Define the intermediate function
\ben
g(r) = {P \over r^4 h} -\alpha~,
\een
then solve for $\dot{z}$ to obtain
\ben\label{zdot}
\dot{z} = {g(r) \over \sqrt{r^4 +B^2 - r^4 h g(r)^2}}~.
\een
The full profile $z(r)$ is obtained by integration. This cannot be done analytically in general, but for any choice of $B$, $P$ and $\alpha$ the integration of \eqref{zdot} is easily evaluated numerically.

These solutions are completely specified by the integration constant $P$. For any brane profile that enters the black hole horizon, substituting $r=1$ into \eqref{pz} shows that $P$ must vanish since $h\lt(r=r_h=1\rt)=0$, 
\ben
\boxed{\Big.~P=0~{\rm for~solutions~with~support~at~} r=1~\Big.}
\een
In keeping with the literature we call these solutions {\it black hole embeddings}. Since $P=0$, we have $g(r)=-\alpha$. Thus, we see from \eqref{zdot} that for the black hole embeddings $z(r)$ is single-valued and monotonic.

The other possibility is that the profile has a minimum value of $r$.  Without loss of generality, we can choose this minimum to be located at $z=0$. The signal of a minimum would be $\dot{z}$ diverging at some $r=r_0$. This yields the expression for the integration constant
\ben\label{przero}
P = r^4_0 \sqrt{h_0} \lt(\sqrt{1+{B^2\over r^4_0}}{\rm sign}\lt(\dot{z}_0\rt) + \alpha \sqrt{h_0}\rt)~,
\een
where $h_0 = h(r_0)$ and $\dot{z}_0 = \dot{z}(r\to r_0)$. The presence of an absolute minimum requires that the brane bends back up to the boundary. This other leg of the brane will have opposite orientation parameter $\zeta$ so this solution is a joined brane/anti-brane pair. We thus call the $P\ne0$ solutions {\it joined embeddings}. 

\begin{figure}[ht]
\centering
\includegraphics[width=.6\textwidth]{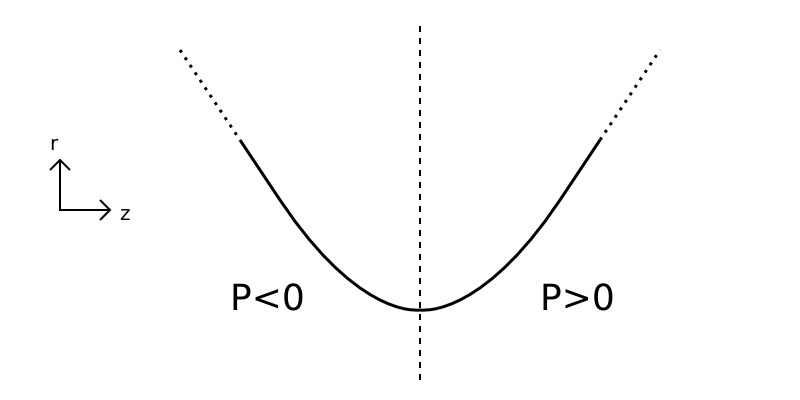}
\caption{\label{minfig} The sign of the integration constant $P$ is the same as that of $\dot{z}$ as $r_0$ is approached and flips accordingly as the minimum at $z=0$ is crossed.}
\end{figure}

The magnitude of the first term in the parentheses of \eqref{przero} is greater than unity while that of the second term is less than unity. Therefore
\ben\label{slopemin}
{\rm sign} \lt(\dot{z}\lt(r\to r_0\rt)\rt)={\rm sign}\lt(P\rt)~.
\een
However, $\dot{z}\to-\infty$ when approaching from the left of the minimum while $\dot{z}\to+\infty$ when approaching from the right. Furthermore, the orientation parameter $\zeta$ changes sign from one branch to the other. Therefore, $P$ changes sign as well,\foot{The reader may find this disconcerting, since $P$ is playing the role of a conserved quantity. The resolution lies in the multi-valuedness of the function $z(r)$. $P$ need only be constant on a given single-valued branch. The minimum is precisely where the single-valued parameterization $z(r)$ breaks down and so consequently does the definition of $P$. That the magnitude of $P$ is constant follows from the continuity of the embedding. } with $P<0$ for $z<0$ and $P>0$ for $z>0$ (see  \figref{minfig}). The joined configuration  is clearly symmetric under parity $z\to -z$, so we can without loss of generality focus our attention to a single branch. We will therefore restrict our attention to $P\ge 0$, which includes the black hole embedding and the ``right branch'' with $\dot{z}_0>0$ of the joined solutions. 

\begin{figure}[ht]
\centering
\begin{tabular}{cc}
\includegraphics[width=.5\textwidth]{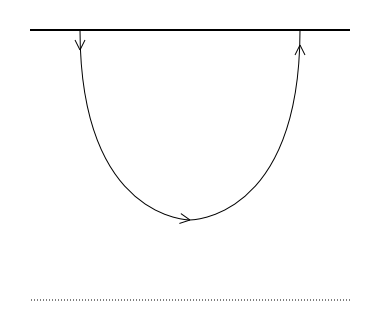} &
\includegraphics[width=.5\textwidth]{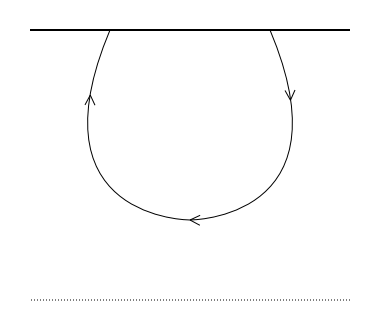}
\end{tabular}
\caption{\label{skinnyfat} ``Skinny'' and ``chubby'' joined embeddings.}
\end{figure}

At the boundary
\ben\label{slopebound}
{\rm sign} \lt(\dot{z}\lt(r\to\infty\rt)\rt) = - {\rm sign} \lt(\alpha\rt)~,
\een
indicating that the direction in which the brane bends initially on its descent from infinity is given entirely by the sign of the $D3$-brane charge. Comparing \eqref{slopemin} and \eqref{slopebound} we see there are thus two qualitative classes of joined solutions, given by the relative sign of $P$ and $\alpha$. For ${\rm sign}(P)= - {\rm sign}(\alpha)$, the sign of $\dot{z}$ remains the same throughout the branch, {\it i.e.} each branch of the brane is separately monotonic. On the other hand, for ${\rm sign}(P)= {\rm sign}(\alpha)$ even a given branch is not monotonic. We call these two possibilities ``skinny'' and ``chubby,'' respectively. See \figref{skinnyfat}.

Physically, we know that the brane and anti-brane have an attraction due to exchange of gravitons and Ramond-Ramond quanta. Further, the background $F_5$ also deflects branes and anti-branes in opposite directions. In the skinny solutions, the background $F_5$ pushes the two stacks together while in the chubby solutions the Ramond-Ramond field forces them apart. 

\subsection{Asymptotics}

The asymptotic separation in $z$ of a joined brane/anti-brane pair is not independent of $r_0$. Define $L$ as
\ben\label{ldef}
L(r_0) = 2\int_{r_0}^\infty \dot{z}(r)~,
\een
where the factor of two arises since the integral is only over one branch of the brane system. For a joined solution, {\it i.e.} any solution with $r_0>1$, $L$ is the asymptotic separation in the $z$ direction of the two ends of the solution. For $r_0=1$ however, the brane and anti-brane are disconnected black hole embeddings. In this case the asymptotic separation is truly a free parameter and $L(r_0=1)$ simply records (twice) the range in $z$ that each branch of the embedding spans.

The probe branes for generic $\alpha$ have the large $r$ behavior
\bea
\dot{z}\lt(r\gg1\rt) = -{\alpha \over \sqrt{1-\alpha^2}} {1\over r^2} + O\lt({1\over r^6}\rt)~,\quad\quad \lt(|\alpha|<1\rt)~.
\eea
The case $|\alpha|=1$ is non-generic. Indeed, expanding \eqref{zdot} in yields
\ben
\dot{z}\lt(r\gg1\rt) = -{\alpha\over \sqrt{1+B^2 +2\alpha P}}+ O\lt({1\over r^4}\rt)~,\quad\quad \lt(|\alpha|=1\rt)~.
\een
It follows that $L$ converges for $|\alpha|<1$ and diverges for $|\alpha|=1$, which means that $|\alpha|=1$ branes ({\it i.e.} $D3$-brane probes) do not intersect the $AdS$ boundary at finite $z$ while those with generic $\alpha$ do. The impossibility of the $D3$ probe to intersect the $AdS$ boundary at finite $z$ may be a symptom of the open string tachyon present at weak coupling at the $(2+1)$-dimensional intersection of $D3$-branes.\foot{Since such a system has $\#ND=2$. See \cite{Polchinski:1998rr}.} Whatever the explanation, we will now restrict our attention to $D5$-branes and $D7$-branes so that we can study probes which intersect the boundary at a finite location.

\begin{figure}[ht]
\centering
\includegraphics[width=.8\textwidth]{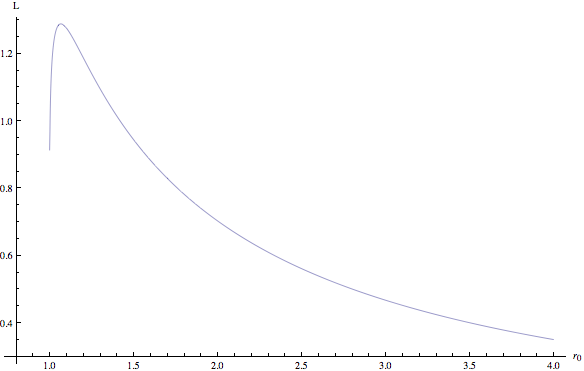}
\caption{\label{lvrsample} Asymptotic brane separation $L$ of a joined solution ($\alpha=-{1\over3}$ and no magnetic field) as a function of minimum radius $r_0$.}
\end{figure}

The right-hand side of \eqref{ldef} is a complicated function of $r_0$ since $\dot{z}$ depends on it through the integration constant $P$. We do not have an analytic expression but can plot it numerically. As an example, see \figref{lvrsample}, which plots $L(r_0)$ for a $D7$-brane probe with $B=0$ and $f_1=f_2={1\over\sqrt{2}}$. Note that $L(r_0)$ is not monotonic and has a maximum. Therefore, when the brane/anti-brane pair are sufficiently separated at the boundary (with an $L\gtrsim 1.3$) there are no joined solutions, only black hole-embeddings. Further, due to the maximum there is a range of $L$ where there are two $r_0$, that is two solutions with the same boundary condition. 

Another feature worth noting is the abrupt end of the curve at $r_0=1$. The $r_0=1$ solution is a black-hole embedding and $L(1)$ is (twice) the $\Delta z$ spanned by a single branch of that embedding. The curve $L(r_0)$ does not continue past this point.

\begin{figure}[ht]
\centering
\includegraphics[width=.6\textwidth]{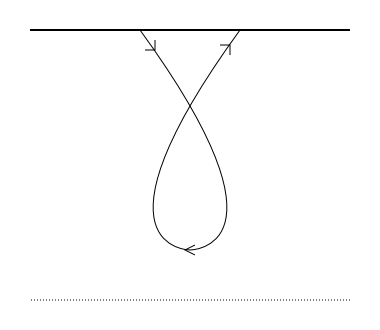}
\caption{\label{fish} An unphysical solution with negative $L$.}
\end{figure}

There is a class of unphysical solutions lurking within the family that we have been discussing. Some of the non-monotonic branches, {\it i.e. } those with ${\rm sign} \lt(P\rt) = {\rm sign} \lt(\alpha\rt)$, will turn out to have negative $L$. Qualitatively these solutions appear as in \figref{fish}. Note that they have the same boundary conditions as a ``skinny'' solution. These solutions are clearly unstable to brane reconnection at the intersection point and will not be considered further.

\section{Free energy}

Now that we have classified the solutions, we investigate the phases of a pair of brane/anti-brane probes. The dynamical problem is to find the solution in a given ensemble, with given boundary conditions, which has the lowest free energy. This solution will dominate and be thermodynamically stable. In the present case, the boundary conditions are given by the asymptotic brane positions and orientations and the values of the fluxes, including magnetic field. Without loss of generality, we can assume\foot{Due to translation invariance in the $z$ direction.} that the center of the pair is at $z=0$, {\it i.e.} if they join, they join at $z=0$. Then the boundary conditions are given by $L$, $B$, and $\alpha$. 

The free energy is conventionally given as the negative of the on-shell action. This is, up to a positive constant, simply the effective action \eqref{efflag} 
\ben\label{fdef}
F (r_0)=\int_{r_0}^\infty dr \lt\{ \sqrt{\lt(r^4 +B^2\rt)\lt(1+r^4 h \dot{z}^2\rt)}+\alpha r^4 h \dot{z}\rt\}~,
\een
This is the free energy of a single leg of the brane/anti-brane system. In the case of $r_0=1$, \eqref{fdef} is the energy of one entire worldvolume, from horizon to boundary. For $r_0>1$, it computes the free energy of one half of the joined brane/anti-brane system. In all cases since the other branch is obtained by symmetry, the true free energy is just twice \eqref{fdef}. Substituting in the general solution \eqref{zdot} we get 
\bea\label{evalfree}
F(r_0) &=& \int_{r_0}^\infty dr \sqrt{r^4+B^2 \over 1 - {r^4 \over r^4+B^2} h g^2} \lt(1+ {r^4\over r^4+B^2} \alpha h g\rt)~,\cr
&=& \int_0^{1\over r_0} {du\over u^4} { 1+B^2 u^4+\alpha h\lt({1\over u}\rt) g\lt({1\over u}\rt)\over  \sqrt{1+B^2 u^4-h\lt({1\over u}\rt) g\lt({1\over u}\rt)^2 } }~,
\eea
where we changed integration variables to $u=r^{-1}$ in the second line. 

Note that \eqref{evalfree} is generically infinite. Indeed, placing a cut-off at the lower end of the $u$ integral yields
\ben
F (r_0) = \int_\epsilon  {du\over u^4} { 1+B^2 u^4+\alpha h\lt({1\over u}\rt) g\lt({1\over u}\rt)\over  \sqrt{1+B^2 u^4-h\lt({1\over u}\rt) g\lt({1\over u}\rt)^2 } } \sim {\sqrt{1-\alpha^2}\over3\epsilon^3}+{\rm finite}~.
\een
Since this divergence is independent of $r_0$, the difference in free energy between any two embeddings will be finite and numerically computable. We will thus compute a renormalized free energy
\ben
\Delta F (r_0) \equiv F(r_0) - F_0~,
\een
with $F_0$ the divergent free energy of the black hole embedding with $r_0=1$,
\ben
F_0 = \int_0^{1} {du\over u^4} \sqrt{1+B^2 u^4-\alpha^2 h\lt({1\over u}\rt)  } ~.
\een
When $\Delta F<0$, the joined solution has less energy than the black hole embeddings and so it dominates, indicating flavor symmetry breaking in the bi-layer description.

\begin{figure}
\centering
\includegraphics[width=.7\textwidth]{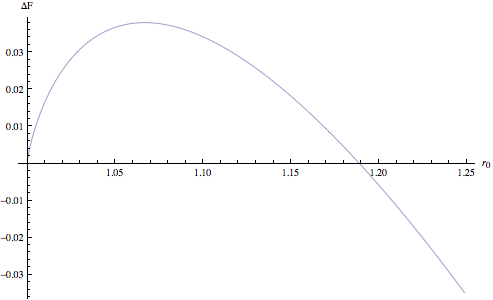}
\caption{\label{FvRsample} The renormalized free energy as a function of $r_0$ for $B=0$ and $\alpha=-{1\over3}$.}
\end{figure}

In \figref{FvRsample} we plot the renormalized free energy as a function of $r_0$ for the case $B=0$ and $\alpha=-{1\over3}$. This is the same set of solutions whose asymptotic separation versus $r_0$ is plotted in \figref{lvrsample}. The only joined solutions with negative free energy are those with $r_0 \gtrsim 1.19$ which corresponds to $L \lesssim 1.2$. For any larger $L$, the black hole embedding is less energetic or the joined embedding does not exist.

\section{Phase diagram and discussion}

\begin{figure}
\centering
\includegraphics[width=1\textwidth]{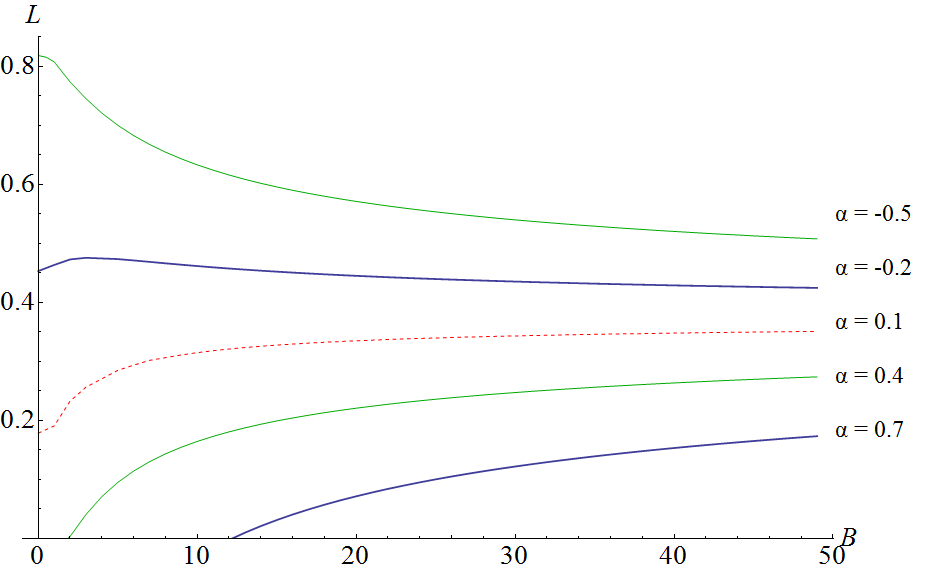}
\caption{\label{LvB} Phase diagrams for the defect system. Above any fixed $\alpha$ curve, the dominant solution is given by the two disconnected brane worldvolumes, {\it i.e.} the symmetric phase. The joined solutions, the broken symmetry phase, dominates below the curve. }
\end{figure}

In \figref{LvB}, we plot the phase diagram of the brane/anti-brane system in the $L$-$B$ plane. Each curve is at fixed $\alpha$, above the curve being the flavor symmetric phase where the stacks do not join while below the curve the symmetry is broken to the diagonal subgroup by brane recombination. We can see that for $\alpha$ negative, the stacks always join at small enough $L$. This is quite intuitive since the background $F_5$ assists the native attraction of the brane and anti-brane so there is no effect to prevent their joining. On the other hand, we see that for large enough positive $\alpha$, the stacks do not join at small $L$ unless there is also a strong enough external magnetic field. Intuitively, the force from the background $F_5$ is strong enough to overcome the brane/anti-brane attraction even at arbitrarily small separation.

In these types of studies, there is a general expectation of {\it magnetic catalysis}, that an external magnetic field favors the breaking of flavor symmetry. This effect has been seen both in perturbative and large-$N$ calculations in quantum field theory \cite{Gusynin:1994re}. It is also known to be a common feature in holographic scenarios of various dimension brane intersections \cite{Johnson:2008vna,Filev:2009xp,Filev:2010pm}. However, in \cite{Preis:2010cq} the Sakai-Sugimoto model was studied at finite chemical potential and magnetic field and an inverse magnetic catalysis was found in a certain region of the phase diagram, {\it i.e.} at zero temperature and fixed finite chemical potential, an increase in magnetic field can prompt a transition to a symmetric state. 

We see in \figref{LvB} both catalysis and inverse catalysis, depending on the value of $\alpha$ and the region of the curve in question. One can see that all positive $\alpha$ embeddings exhibit catalysis, {\it i.e.} all of the chubby solutions. In these cases, it appears that external magnetic field always enhances the attraction of the brane/anti-brane pair. However, for $0 \gtrsim \alpha \gtrsim -.2$ the curves are similar to positive $\alpha$ so the sign of the induced $D3$ charge is not sufficient to determine the behavior with respect to magnetic field. For $\alpha\approx -.2$, we see a maximum, indicating a region of inverse catalysis for small $B$. This region expands as $\alpha$ is decreased until there is inverse catalysis for all $B$.

It is not clear from the point of view of the field theory what dictates whether the system exhibits catalysis or inverse catalysis. We will refrain from speculating on the exact mechanism here and leave this question to future work.

\begin{appendix}

\subsection*{Acknowledgments}
 This work is supported in part by the Natural Sciences and Engineering Research
Council of Canada. The authors would like to thank the following individuals for fruitful discussion related to this work: Per Kraus, Thomas Levi, Hamid Omid, Gordon Semenoff, and Mark Van Raamsdonk.

\end{appendix}

\end{document}